\documentclass[aps,preprintnumbers,prd,twocolumn,superscriptaddress,nofootinbib]{revtex4-1}
\usepackage{blindtext}
\usepackage{lineno}
\usepackage{bm,color,xcolor}
\usepackage{slashed} 
\usepackage{graphicx}
\usepackage{subfigure} 
\usepackage{soul} 
\usepackage{multirow}
\usepackage{float}
\usepackage{comment}
\usepackage{fontawesome} 
\usepackage{mathtools}
\usepackage[colorlinks=true
,urlcolor=blue
,anchorcolor=blue
,citecolor=blue 
,filecolor=blue
,linkcolor=blue
,menucolor=blue
,linktocpage=true
,pdfproducer=medialab
,pdfa=true
]{hyperref}
\usepackage{lipsum}

\makeatletter
\def\l@subsubsection#1#2{}
\makeatother

\newcommand{\beq}{\begin{equation}}
\newcommand{\eeq}{\end{equation}}
\newcommand{\bea}{\begin{eqnarray}}
\newcommand{\eea}{\end{eqnarray}}

\newcommand{\brr}[1]{\left(#1\right)}

\usepackage[arrows]{ezedits}
\defineEditQuick{ZL}{red}

\begin{document}


\preprint{UMN-TH-4324/24}
\preprint{FTPI-MINN-24-15}

\title{$W$-Boson Exotic Decay into Three Charged Leptons at the LHC}

\author{Yifan Fei}
\email{20307130006@fudan.edu.cn}
\affiliation{Department of Physics and Center for Field Theory and Particle Physics,
Fudan University, Shanghai 200438, China}
\author{Peiran Li} 
\email{li001800@umn.edu}
\author{Zhen Liu}
\email{zliuphys@umn.edu}
\author{Kun-Feng Lyu} 
\email{lyu00145@umn.edu}
\affiliation{School of Physics and Astronomy, University of Minnesota, Minneapolis, MN 55455, USA}
\author{Maxim Pospelov}
\email{pospelov@umn.edu}
\affiliation{School of Physics and Astronomy, University of Minnesota, Minneapolis, MN 55455, USA}
\affiliation{William I. Fine Theoretical Physics Institute, School of Physics and Astronomy,
University of Minnesota, Minneapolis, MN 55455, USA}

\begin{abstract}
We investigate the $W$ boson's exotic decay channel, $W \rightarrow \ell\ell\ell \nu$, at the LHC. Although the four-body final states suppress the decay branching ratio, the large production of $W$ bosons makes detecting and precisely measuring this decay probability entirely feasible. 
Our simulation study indicates that this tiny branching ratio can be measured with sub-percent precision at the HL-LHC. This decay channel can also constrain Standard Model extensions. Using the $ L_\mu-L_\tau$ model as a benchmark, we find that the current bound on the gauge coupling for $Z'$ mass in the range of $[4,75]$ GeV can significantly improve. 
\end{abstract}

\maketitle

\setcounter{secnumdepth}{3}
\setcounter{tocdepth}{1}
\tableofcontents

\section{Introduction}
Precision measurements of the Standard Model (SM) are a powerful method to test its internal consistency and explore new physics, i.e., possible Beyond SM (BSM) extensions. $W$ bosons, predicted as the minimal UV-completion of the four-Fermi weak interactions, were discovered over forty years ago. One might think that many decay channels of $W$ bosons have been studied by now. However, only a limited number of decay modes have been measured thus far. The Large Hadron Collider (LHC) has been extremely successful as a Higgs machine, a top quark machine, and a direct BSM search machine. Future plans aim to extend the data sample to an unprecedented integrated luminosity of 3~ab$^{-1}$~\cite{ZurbanoFernandez:2020cco}.
We argue that it will also produce and record the highest number of $W$ bosons~\cite{ATLAS:2018wis,ParticleDataGroup:2022pth,Salvatico:2021zky,ATLAS:2023jfq} in accelerators, $N_W \sim O(10^{11})$, and hence can reveal new physics through the $W$ boson's rare decays. Specifically, we show that the LHC has collected enough data to provide a few percent level precision measurement on a rare $W$ boson decay~\cite{CDF:2011jwr,CMS:2020oqe,Mangano:2014xta,CMS:2019vaj} within the SM through the radiative process. Additionally, this channel can probe various interesting new physics, e.g., HNL searches~\cite{ATLAS:2019kpx,ATLAS:2022atq,Cvetic:2018elt} with competitive sensitivity. Our paper should be considered complementary to (and independent from) the recent study by ATLAS of the exotic $Z'$ searches via the $W\rightarrow 3\ell+\nu$ mode~\cite{ATLAS:2024uvu}.

The paper is organized as follows: First, we discuss the SM expectation on the rare, radiative $W$ decay, $W\rightarrow 3\ell+\nu$, as a function of the two charged lepton invariant mass threshold. Subsequently, the decay branching ratio measurement at the LHC is investigated. We define the fiducial branching ratio to tackle the contribution from other diagrams. Furthermore, this rare decay channel is utilized to impose constraints on the parameter space of the $L_\mu - L_\tau$ model. Finally, we provide a summary and outlook.

\section{(On-shell) Decay Width from Theory}
\label{sec:Br_compute}
\begin{figure}[hb]
    \centering
    \includegraphics[width=8.0cm]{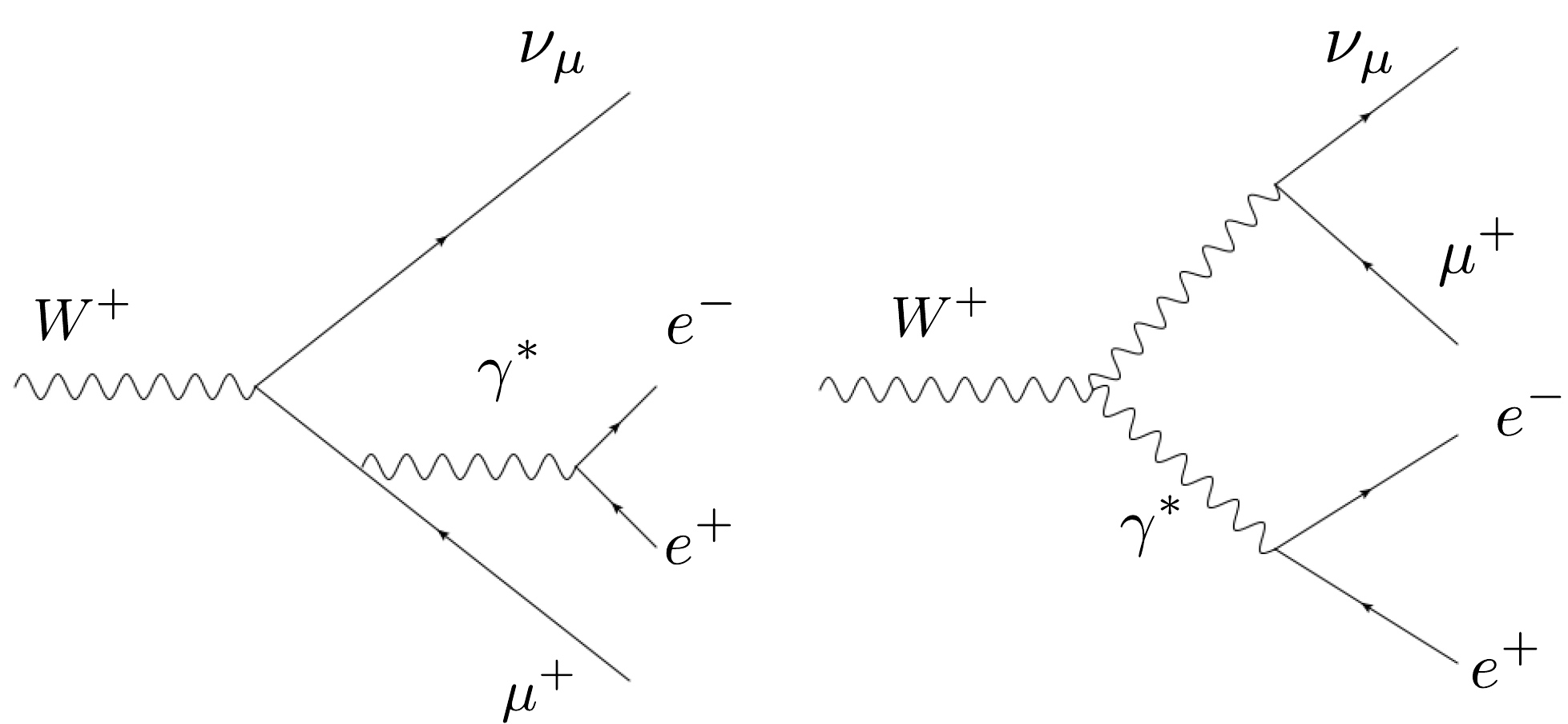}
    \caption{The dominant Feynman diagrams for the decay process $W^+ \rightarrow \mu^+ \nu_\mu e^+ e^-$.}
    \label{fig:feynman_diagram}
\end{figure}

We first compute the theoretical prediction for the on-shell decay branching ratio. In the SM, multiple Feynman diagrams contribute to the 1-to-4 decay process $W^+ \to \mu^+ \nu_\mu \ell^+ \ell^-$, where $\ell$ denotes electron or muon. The dominant contribution is from the two diagrams shown in \autoref{fig:feynman_diagram}. The $W^+$ boson decays into an anti-muon and a muon neutrino, followed by a virtual photon converting to a pair of electron and positron. 
The off-shell photon can originate from either the muon or the $W$ boson. The invariant mass of the electron pair determines the off-shellness of the intermediate photon. If the electron pair is too soft or nearly collinear with the decaying $W$, the photon propagator would approach the pole, requiring further resummation. Practically, to measure the final three charged leptons, we need to apply kinematic cuts on the electron pair to make them visible and separable. Hence, we impose a threshold on the electron pair invariant mass and estimate the decay branching ratio. The partial decay width for the process $W^+(p) \rightarrow \mu^+(p_1) \nu_\mu(p_2) e^+(k_1) e^-(k_2)$ is given by integrating over the phase space of the four final particles.

\begin{figure}[th]
    \centering
    \includegraphics[width=8cm]{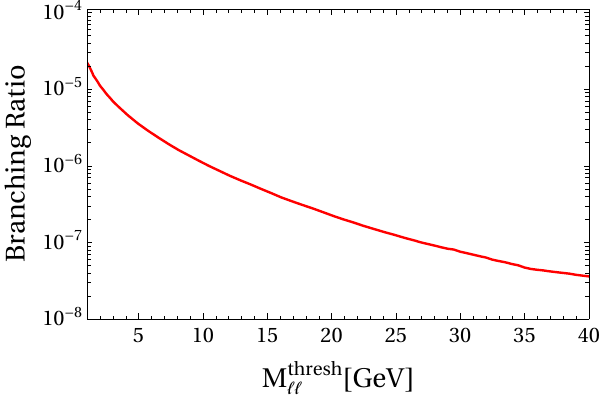}
    \caption{The branching ratio of the decay channel $W^+ \rightarrow \ell^+ \ell^-\ell^+ \nu_\ell$ as a function of the charged lepton pair invariant mass $\text{M}^{\rm thresh}_{\ell\ell}$ threshold value.}
    \label{fig:W_decay_Br_fun}
\end{figure}

We use the package {\tt FeynCalc}~\cite{Shtabovenko:2023idz,Shtabovenko:2020gxv} to perform the numerical integration over the phase space region. As a cross-check, the package {\tt MadGraph5\_aMC@NLO}~\cite{Alwall:2014hca} is used to generate the decay sample files, and the invariant mass cuts are imposed to obtain the corresponding particle decay branching ratio. The only cut is on the invariant mass of the electron pair. We set its invariant mass $\text{M}_{\ell\ell}$ larger than $\text{M}^{\rm thresh}_{\ell\ell}$ and then scan this threshold parameter. In \autoref{fig:W_decay_Br_fun}, we plot the decay branching ratio as a function of $\text{M}^{\rm thresh}_{\ell\ell}$. The invariant mass starts from 1 GeV, at which the branching ratio is as high as $O(10^{-5})$. As the cut value increases, the amplitude becomes more suppressed due to the photon propagator. 

Note that in this calculation the initial $W$ boson is exactly on-shell. However, due to the large decay width of the $W$ boson, the real production of the decay process cannot be factorized exactly. In the next section, we will first define the fiducial decay branching ratio and compare it with the on-shell decay width we have computed.

\section{Observable Fiducial Decay Width}
\label{sec:Br_measurement}
In this section, we focus on the measurable exotic decay branching ratio of the $W$ boson in the SM. 
The expected signal final states are the three charged leptons from the $W$ boson decay. 
Under the zero-width approximation, the signal cross section can be factorized into the product of the $W$ boson production rate and the decay branching ratio, namely
$\sigma(p p \rightarrow W) \times \text{Br}(W \rightarrow \ell \ell \ell \nu_\ell)$. 
The produced $3\ell\nu_\ell$ invariant mass is not highly peaked at $m_W$ but rather has a broader extension with long tails. 
Moreover, additional diagrams contribute to the channel $p p \rightarrow \ell \ell \ell \nu_\ell$. 

To address these subtleties, a clear definition should be made to separate the signal from the background. Because a slightly off-shell $W$ boson can turn into an on-shell $W$ boson with a virtual photon (similar to the right pannel of \autoref{fig:feynman_diagram}), the signal must be defined in a narrow window, unlike the inclusive $W/Z$ boson measurements~\cite{CMS-PAS-SMP-20-004,ATLAS:2016fij}. Any event with the four-lepton invariant mass within the On-shell Region (OR) 
\begin{equation}
    m_W - 2 \Gamma_W < \sqrt{\hat s} < m_W + 2 \Gamma_W,
\end{equation}
is counted as the signal. Events outside this range are considered background.
At the parton level, this accounts for most of the $W$ bosons near the pole, making the rate defined this way unambiguous. It is a typical subtlety in attempts to extract decay branching fractions of unstable particles, where one must compare and convert the convoluted rate around the pole region to an on-shell quantity. One can choose to enlarge the $\pm2\Gamma_W$ region when defining the signal region at the parton level, and the result will be robust against such choices as long as signal and background are treated consistently during the simulation process. 
We aim to extract the branching ratio dominantly from the $W$ decay and avoid contamination from other $t$-channel diagrams. The fiducial branching ratio can be extracted by
\begin{align}
    \text{Br}^{\text{fid}} = \frac{\sigma^{\text{fid}}_{\text{signal}}}{\sigma_W} \times f_W,
\end{align}
where $\sigma^{\text{fid}}_{\text{signal}}$ is the signal cross section after required kinematic selections at the reconstruction level, $\sigma_W$ is the inclusive $W$ boson cross section, and $f_W$ is the theoretical ratio of the fiducial cross section of $pp \to W \to \ell\ell\ell\nu$ to our fiducial signal cross section:
\begin{align}
    f_W = \frac{\sigma(p p \rightarrow W) \times \text{Br}(W \rightarrow \ell \ell \ell \nu_\ell)}{\sigma^{\text{fid}}_{\text{signal}}}.
\end{align}
For consistency, we generate the on-shell decay sample files in the rest frame of the $W$ boson and then boost them to the lab frame, imposing the same cuts as the fiducial signal. $f_W$ is found to be larger than 1 due to the negative interference between the $s$-channel and $t$-channel diagrams. 
Such negative interference will cause a smaller signal cross section, resulting in a slightly larger $f_W$. 
It will also decrease if the OR is widened. Nevertheless, we aim to estimate the precision $\frac{\delta \text{Br}}{\text{Br}}$ which is independent of $f_W$. This is another interesting example of non-factorizable SM on-shell interference~\cite{Campbell:2017rke}; it is related to the radiation zeros~\cite{Mikaelian:1979nr,Baur:1994ia,Han:1995ef,Baur:1999ym} and warrants further theory explorations.

\section{$W$ Boson Exotic Decay Precision Measurement}
\label{sec:Br_measurement}

After defining the observable fidual branching fraction, we study the projected sensitivity. We choose two benchmark collider configurations. 
The primary benchmark is based on the future 14~TeV HL-LHC~\cite{Apollinari:2015bam} with an integrated luminosity of $3~\text{ab}^{-1}$. 
The second benchmark implements a conservative analysis of the 13~TeV Run-2 stage following CMS/ATLAS configurations with an integrated luminosity of $150~\text{fb}^{-1}$.
Before enumerating all possible signals and backgrounds, we highlight that the signal process exhibits a distinct characteristic of three leptons in the final state with a low invariant mass:
\begin{equation}
    \text{M}_{\ell\ell\ell}<80~\text{GeV}.
\end{equation}
This unique characteristic allows us to discriminate this rare decay from the majority of multi-lepton background.

Next, we classify the signal and background channels. The parton-level events are generated via the package {\tt MadGraph5\_aMC@NLO}. They are subsequently passed to the interfaced {\tt Pythia8}~\cite{Bierlich:2022pfr} and {\tt Delphes3}~\cite{deFavereau:2013fsa,Mertens:2015kba} to simulate the reconstruction-level events under the two benchmark detector configurations. 
Various kinematic selections and deep neural network (DNN) analyses are implemented to extract the final precision of the branching ratio measurement.

\subsection{Parton Level Treatment}

\subsubsection*{Main Signal}
There have been dedicated measurements of $W$ boson production at the LHC~\cite{ATLAS:2016nqi,ATLAS:2019fgb,ATLAS:2016fij,ATLAS:2017rzl,CMS:2011aa,CMS:2015tmu,CMS:2023foa,CMS:2020cph}. Theoretical predictions of the total cross section involve NNLO QCD corrections~\cite{Melnikov:2006di,Anastasiou:2003ds,Anastasiou:2003yy} plus complete NLO electroweak corrections~\cite{Dittmaier:2001ay,Baur:1998kt,Gavin:2012sy}. The dominant systematic uncertainty stems from the proton PDFs.
We generate the parton-level events at the tree level and then perform jet matching to resolve the infrared divergence. 
The leading order channel for the on-shell $W$ boson production is the 2-to-1 process: $u + \bar{d} \rightarrow W^+$ with its conjugate process. Additionally, the associated multiple jets production is generated to account for the QCD correction. Overall, we run the following processes in {\tt MadGraph5\_aMC@NLO} at the parton level: 
\[pp \to \ell\ell\ell\nu_\ell +(j),~\text{M}_{\ell\ell\ell\nu}\in \text{OR}\]
where $\ell$ denotes $e$ and $\mu$, and $j$ refers to the jet including gluons and light quarks. The $\tau$ lepton also contributes to this channel; however, the charged-lepton from $\tau$ decay would be too soft to pass the trigger, and hence, its contribution is negligible. One can study a $\tau$-specific exotic decay using various hadronic $\tau$-taggers~\cite{Furukawa:2024dsh,Ordek:2019vrf,CMS:2018jrd,CMS:2020cmk}. Therefore, the main contribution is from the $e,\mu$ final states. The parton-level generator cut, mainly for the purpose of improving signal generation efficiency, for the HL-LHC (Run-2) benchmark is
\begin{align*}
    &p_T(j)>20~\text{GeV},~p_T(\ell)>3~\text{GeV},~\eta(\ell)<5.0(2.5), \\
    &~\eta(j)<5.0,~\Delta R(\ell\ell)>0.2,~\text{M}_{\ell\ell}>1~\text{GeV},
\end{align*}
where $\text{M}_{\ell\ell}$ is the invariant mass of the dilepton.
The signal events then pass through the interfaced Pythia for parton shower and jet matching. 
We match up to one jet and choose the merging scale to be $15~$GeV.
After jet matching, the cross section is shown in \autoref{Table:parton_level_signal_xs}.
\begin{table}[th]
\centering
\begin{tabular}{|c|c|}
\hline
Signal process & Cross section [pb] \\
\hline
$pp \to \ell^+ \ell^- \ell^+ \nu_\ell + (j),~\text{M}_{\ell\ell\ell\nu}\in \text{OR}$ & $0.36$  \\
\hline
$pp \to \ell^+ \ell^- \ell^- \Bar{\nu}_\ell + (j),~\text{M}_{\ell\ell\ell\nu}\in \text{OR}$ & $0.25$  \\
\hline 
\end{tabular}%
\caption{Cross section of the signal production after the parton-level initial selection and jet merging within the HL-LHC benchmark, dominated by the on-shell $W^+$ and $W^-$ decays.}
\label{Table:parton_level_signal_xs}
\end{table}

\subsubsection*{Background Considerations}
The primary backgrounds can be classified as
\begin{enumerate}
    \item $p p \to \ell \ell \ell \nu_\ell + (j),~\text{M}_{\ell\ell\ell\nu}\notin \text{OR}$
    \item $p p \to \ell \ell \ell \ell + (j)$
    \item $p p \to t \bar{t} + (j)$.
\end{enumerate}

The first class of background, $p p \to \ell \ell \ell \nu_\ell + (j),~\text{M}_{\ell\ell\ell\nu}\notin \text{OR}$, involves three charged leptons plus one missing neutrino final state with the phase space outside the OR regime. 
The four-lepton invariant mass can be either below $m_W - 2 \Gamma_W$ or above $m_W + 2 \Gamma_W$. 
In the low invariant mass region, the four-lepton final states can come from various off-shell states or low-lying hadron decays and are difficult to veto.
The events with the lepton invariant masses greater than $m_W+2\Gamma_W$ are dominated by the on-shell $W$ and $Z$ boson production with leptonic decay, typically having $\text{M}_{\ell\ell\ell}>m_Z$.

For the second class, $p p \to \ell \ell \ell \ell + (j)$, we consider the four charged leptons final states with one lepton lost at the reconstruction level. One of the primary contributions is the single $Z$ boson exotic decay, similar to what we are focusing on~\cite{ATLAS:2021kog,CMS:2017dzg}. Both the single $Z$ boson production rate and its exotic decay branching ratio are smaller than the $W$ boson by a factor of a few. 
Other principal channels include the production of a double $Z$ boson or a single $Z$ boson with a virtual photon, followed by decay into two pairs of leptons. 

The $t\bar{t}$ production is also included as part of the main backgrounds. At $\sqrt{s} = 13$ TeV, the cross section of top quark pair production is around 800 pb. A top quark can decay into a bottom quark plus $W$ boson followed by leptonic decay final states. The $b$-jets can be further misidentified as an isolated lepton. As a result, we may observe two real leptons plus one faked ``lepton" at the reconstruction level. Although the mis-tag rate is roughly $10^{-3}$~\cite{ATLAS:Fake_lepton}, it cannot be ignored due to the huge production rate.

All backgrounds are implemented with the same parton-level pre-selection as the signal events. We also include jet matching and choose the same merging scale as the signal. After merging, the cross section for each background process is listed in the second column of \autoref{Table:cutflow_table_HLLHC} and \autoref{Table:cutflow_table_run2}.

\subsection{Reconstruction Level Analysis}

After passing through {\tt Pythia8} for parton shower and jet merging, the truth-level sample is sent to {\tt Delphes3} for fast parametric simulation to include various detector effects. We carry out two comparative analyses: one is cut-based, and the other is DNN-based. A reconstruction-level initial selection according to both the current trigger system from ATLAS and CMS~\cite{CMS:2021yvr,ATLAS:2020gty} and future HL-LHC detector performance~\cite{Ryd:2020ear,Guiducci:2018ogp} is shown in this section.

We show the invariant mass of the three charged leptons in \autoref{fig:W_decay_Br_fun}. The three charged leptons from the $W$ boson tend to share half of the total invariant mass due to the favored-low invariant mass of two same-flavor charged leptons from the fermion propagator suppression. Hence, keeping the soft charged leptons is key to sizable signal statistics. 
In the following, we show two analyses based on future and current multi-lepton triggers and the projected sensitivity for HL-LHC and LHC Run-2, respectively.


\subsubsection*{HL-LHC Analysis}
The multi-lepton requirement primarily determines the signal efficiency. According to the current lepton trigger system~\cite{CMS:2021yvr,ATLAS:2020gty}, the lower threshold of transverse momentum is $4~\text{GeV}$, and we assume we can keep the low threshold at HL-LHC. The rapidity coverage will be improved in the future HL-LHC~\cite{Ryd:2020ear,Guiducci:2018ogp}.
Thus, the HL-LHC benchmark is treated under an initial selection of
\begin{align*}
    &\eta(\mu)<2.8,~\eta(e,j)<4,~\Delta R(\ell\ell,j\ell)>0.2,\\
    &n(\ell)=3, ~p_T(\ell)>5~\text{GeV},\\
    &n(j)\leq 2~\text{with}~p_T(j)>20~\text{GeV},~ \text{M}_{\ell\ell}>4~\text{GeV}
\end{align*}
where $n(j)$ and $n(\ell)$ stand for the number of jets and charged leptons. The additional jet veto is to reject the $t\bar{t}$ background because hadronic processes generally produce more jets. The invariant mass of the same-flavor, opposite-charged lepton pair is required to be sizable to avoid the low mass meson resonance. Future detailed analysis could extend the search to this low invariant mass regime.

\begin{widetext}

\begin{table}[ht]
\centering
\begin{tabular}{|c|c|c|c|c||c|c|}
\hline
 \multirow{2}{*}{Cross section $[\text{pb}]$} & \multirow{2}{*}{Parton-level} & \multirow{2}{*}{$n(\ell)=3$} & $n(j)\leq 2,$ & \multirow{2}{*}{$\text{M}_{\ell\ell\ell}<80~\text{GeV}$} & Cut-based result & DNN-based result\\
 \cline{6-7}
 & & & $\text{M}_{\ell\ell}>4~\text{GeV}$ & & $\text{M}_{\ell\ell\ell}<60~\text{GeV}$ & DNN selection\\
\hline
Signal & $0.61(100\%)$ & $0.036(5.9\%)$ & $0.021(3.5\%)$ & $0.021(3.5\%)$ & $0.021(3.4\%)$ & $0.017(2.7\%)$\\
\hline
$pp \to \ell\ell\ell\nu,~\text{M}_{\ell\ell\ell\nu}\notin\text{OR}$ & $0.95(100\%)$ & $0.22(23\%)$ & $0.2(21\%)$ & $0.013(1.4\%)$ & $8\times 10^{-3}(0.87\%)$ & $3.3\times 10^{-3}(0.3\%)$\\
\hline
$pp \to \ell\ell\ell\ell$ & $0.34(100\%)$ & $0.068(20\%)$ & $0.061(18\%)$ & $0.017(5\%)$ & $7.2\times 10^{-3}(2.1\%)$ & $3.2\times 10^{-3}(0.95\%)$\\
\hline 
$pp \to t\Bar{t} + (j)$ & $688(100\%)$ & $0.19(0.027\%)$ & $0.11(0.016\%)$ & $0.023(3\times10^{-5})$ & $0.01(1\times10^{-5})$ & $2.1\times 10^{-3}(3\times10^{-6})$\\
\hline 
\end{tabular}
\caption{Cutflow table of signal and backgrounds under 14 TeV HL-LHC benchmark. The $p_T$, $\eta$ and $\Delta R$ cuts are absorbed into $n(\ell)=3$. The machine learning analysis is separated from the cut-based analysis. There is no successive relation between the last two columns.}
\label{Table:cutflow_table_HLLHC}
\end{table}
\end{widetext}

The cutflow table is shown in \autoref{Table:cutflow_table_HLLHC}.  
Requiring $n(\ell)=3$ excludes a large portion of the signal events since there is a higher probability of two soft same-flavor charged leptons appearing. After the initial selection and requiring $\text{M}_{\ell\ell\ell}<80~$GeV, we proceed with a cut-based analysis and Deep Neural Network (DNN)-based machine learning analysis to separate signal and background. The DNN-based analysis details can be found in \autoref{sec:DNN}. The projected sensitivities of these analyses, based on statistical uncertainty, are:
\begin{align}
    \text{Cut-based:}~~& \frac{\delta\text{Br}^{\text{fid}}(W\to\ell\ell\ell\nu_\ell)}{\text{Br}^{\text{fid}}(W\to\ell\ell\ell\nu_\ell)}=0.71\%,\\
    \text{DNN-based:}& \frac{\delta\text{Br}^{\text{fid}}(W\to\ell\ell\ell\nu_\ell)}{\text{Br}^{\text{fid}}(W\to\ell\ell\ell\nu_\ell)}=0.62\%.
\end{align}

\begin{figure}[t!]
    \centering
    \includegraphics[width=0.45\textwidth]{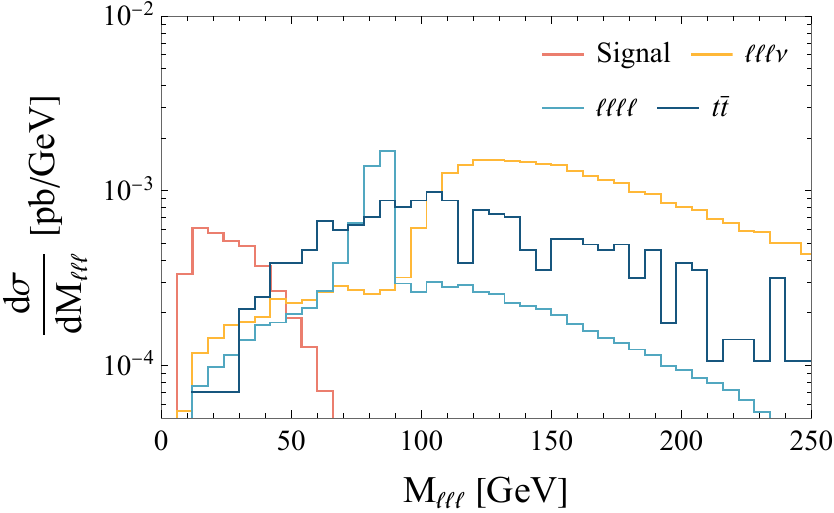}
    \caption{Reconstruction level distribution of $\text{M}_{\ell\ell\ell}$ after initial selection cuts under HL-LHC benchmark.}
    \label{fig:lll_mass}
\end{figure}

As mentioned previously, the invariant mass of the three charged leptons is a strong feature to reject most of the background processes, as seen in \autoref{fig:lll_mass}. Cross-checked with the machine learning results, a simple cut-based analysis is also implemented by only requiring $\text{M}_{\ell\ell\ell}<60~\text{GeV}$, leading to a cut-based analysis precision projection at HL-LHC of 0.71\%. 
The 3-lepton invariant mass $\text{M}_{\ell\ell\ell}$ is peaked at the lower mass regime, as the neutrino tends to carry away half of the energy in the center-of-mass frame. 
The $\ell\ell\ell \nu_\ell$ background is suppressed in the low mass regime. Starting from around 100 GeV, the differential distribution rate sharply increases. 
The on-shell $W$ boson plus $\gamma^*/Z$ starts to play a role.
For the $4\ell$ final states channel, there is a peak at around $m_Z$. This implies that the three $\ell$ are primarily from the single $Z$ boson exotic decay and one soft lepton is invisible. The $\gamma^*/Z$ may not contribute significantly, as the virtual photon tends to be forward. 
Interestingly, the $4\ell$ final state decayed from $Z$ boson is significantly vetoed by the invariant mass selection. For $Z$ boson exotic decay, it will likely miss a soft lepton, which only carries a tiny fraction of the total energy. Consequently, most of the $Z$ boson background will still have an invariant mass larger than 80~GeV. While for $W$ boson exotic decay, the neutrino would take away half of the energy in the center-of-mass frame, hence suppressing the three charged leptons' invariant mass. This is the key distinction between $W$ and $Z$ boson exotic decay.\\

\subsubsection*{Run-2 Analysis}

The 13 TeV Run-2 benchmark is treated more conservatively. Various multi-lepton studies~\cite{CMS:2019lwf,ATLAS:2021wob,ATLAS:2019wgx,ATLAS:2021kog} from CMS/ATLAS set up higher $p_T$ selection to ensure highly efficient triggering of events. Hence, our Run-2 benchmark initial selection is
\begin{align*}
    &\eta(\ell,j)<2.5,~\Delta R(\ell\ell,j\ell)>0.2,~n(\ell)=3,\\
    &p_T^1(\ell) > 20~\text{GeV}, ~p_T^2(\ell) >10~\text{GeV},~p_T(\ell) > 5~\text{GeV},\\
    &n(j)\leq 2~\text{with}~p_T(j)>20~\text{GeV},~ \text{M}_{\ell\ell}>4~\text{GeV}
\end{align*}
where the superscript stands for the leading and sub-leading charged lepton. 

The corresponding cutflow table is shown in \autoref{Table:cutflow_table_run2}. The parton level cross section at the Run-2 benchmark is smaller than that of the HL-LHC due to the different initial selection cuts. In our detector simulation, HL-LHC has a slightly better detection resolution, which results in the $t\bar{t}$ background having a smaller faking probability compared with the Run-2 result. 
Nevertheless, a comprehensive analysis of the faked-lepton background can be handled using data-driven methods by experimental collaborations. 
This analysis only provides an estimation of each process via {\tt Delphes3} fast simulation. Implementing DNN optimization, we end with a precision of
\begin{align}
    \text{Cut-based:}~~& \frac{\delta\text{Br}^{\text{fid}}(W\to\ell\ell\ell\nu_\ell)}{\text{Br}^{\text{fid}}(W\to\ell\ell\ell\nu_\ell)}=6.5\%,\\
    \text{DNN-based:}& \frac{\delta\text{Br}^{\text{fid}}(W\to\ell\ell\ell\nu_\ell)}{\text{Br}^{\text{fid}}(W\to\ell\ell\ell\nu_\ell)}=4.4\%.
\end{align}  
based on statistical uncertainty. 

The precision is comparable with the statistical uncertainty of the $Z\to4\ell$ branching ratio, which is around 3\%~\cite{ATLAS:2021kog,CMS:2017dzg}. In general, the single $W$ boson production rate at the LHC is higher than the $Z$ boson by a factor of 3. Furthermore, the branching ratio of $W$ boson 4-lepton decays should also be larger than that of the $Z$ boson, which can be realized from their dilepton decay. However, we conservatively use $\Delta R(\ell\ell)>0.2$, which significantly reduces signal efficiency. A lower threshold, similar to that in Refs.~\cite{ATLAS:2021kog,CMS:2017dzg}, could further improve the measurement.

We comment on the systematic uncertainties here. One of the major systematics is the $W$ boson production rate from PDFs and higher-order corrections. In the HL-LHC benchmark analysis, the low statistical uncertainty of $0.62\%$ motivates us to reduce the systematic uncertainty of $O(1\%)$ as much as possible.
For instance, the proton PDF uncertainty is around 3\%. However, the measured parameter can be chosen as
\begin{align*}
    \frac{\text{Br}(W\to \ell\ell\ell \nu_\ell)}{\text{Br}(W\to \ell \nu_\ell)},
\end{align*}
which is largely independent of proton PDF choices.

\begin{widetext}
\begin{table*}[t]
\centering
\begin{tabular}{|c|c|c|c|c||c|c|}
\hline
 \multirow{2}{*}{Cross section $[\text{pb}]$} & \multirow{2}{*}{Parton-level} & \multirow{2}{*}{$n(\ell)=3$} & $n(j)\leq 2,$ & \multirow{2}{*}{$\text{M}_{\ell\ell\ell}<80~\text{GeV}$} & Cut-based result & DNN-based result\\
 \cline{6-7}
 & & & $\text{M}_{\ell\ell}>4~\text{GeV}$ & & $\text{M}_{\ell\ell\ell}<60~\text{GeV}$ & DNN selection\\
\hline
Signal & $0.32(100\%)$ & $0.016(5.1\%)$ & $0.011(3.3\%)$ & $0.011(3.3\%)$ & $0.01(3.1\%)$ & $8\times 10^{-3}(2.5\%)$\\
\hline
$pp \to \ell\ell\ell\nu,~\text{M}_{\ell\ell\ell\nu}\notin\text{OR}$ & $0.45(100\%)$ & $0.16(36\%)$ & $0.14(31\%)$ & $9\times 10^{-3}(2\%)$ & $5.6\times 10^{-3}(1.2\%)$ & $1.6\times 10^{-3}(0.36\%)$\\
\hline
$pp \to \ell\ell\ell\ell$ & $0.34(100\%)$ & $0.1(29\%)$ & $0.088(26\%)$ & $0.022(6.4\%)$ & $6.8\times 10^{-3}(2\%)$ & $3.3\times 10^{-3}(0.96\%)$\\
\hline 
$pp \to t\Bar{t} + (j)$ & $584(100\%)$ & $0.26(0.045\%)$ & $0.13(0.022\%)$ & $0.033(0.0056\%)$ & $0.015(0.0026\%)$ & $2\times 10^{-3}(3\times10^{-6})$\\
\hline 
\end{tabular}
\caption{Cutflow table of signal and backgrounds under 13 TeV Run-2 benchmark. The $p_T$, $\eta$ and $\Delta R$ cuts are absorbed into $n(\ell)=3$. The machine learning analysis is separated from the cut-based analysis. There is no successive relation between the last two columns.}
\label{Table:cutflow_table_run2}
\end{table*}
\end{widetext}

\section{Sensitivities in $Z'$ models}
\label{sec:Lmu-Ltau}

In this section, we aim to exploit the $W$ boson exotic decay channels and probe directly BSM models. We first take the $L_{\mu} - L_\tau$ model and then also discuss the results for $Z'$ with anomalous couplings.

The minimal $L_{\mu} - L_\tau$ model~\cite{He:1990pn,He:1991qd,Heeck:2011wj,Langacker:2008yv,Leike:1998wr,Carena:2004xs,Altmannshofer:2014pba} posits that both muon and tau lepton numbers are oppositely charged. It can be spontaneously broken, leading to a massive $Z'$ boson. 
In the IR, this scenario can be parameterized in terms of the gauge coupling $g_{Z'}$ and the mass $m_{Z'}$. 
The $Z'$ boson can be radiated from both of the lepton legs from the 2-body leptonic decays of the $W$ boson.
The relevant Lagrangian is given by
\begin{align*}
    \mathcal{L} &\supset -\dfrac{1}{4} F'_{\mu\nu} F^{\prime \mu\nu} + \dfrac{1}{2} m_{Z'}^2 Z^{\prime 2}_\mu 
    + g_{Z'} Z'_\nu \brr{\bar{\mu} \gamma^\nu \mu - \bar{\tau} \gamma^\nu \tau}\\
    &\quad+g_{Z'} Z'_\nu \brr{\bar{\nu}_\mu \gamma^\nu P_L \nu_\mu - \bar{\nu}_\tau \gamma^\nu P_L \nu_\tau}.
\end{align*}
The signal event is chosen to be the trimuon final state:
\begin{equation}
    pp\to W^{(*)}+(j) \to Z'\mu\nu_\mu +(j) \to \mu\mu\mu\nu_\mu +(j).
\end{equation}
Similar to before, $\tau$-flavored events have very soft leptons with low tagging efficiency and hence require a separate study to optimize a search. We hence focus on the tri-muon final state here. 
As shown in \autoref{Table:cutflow_table_HLLHC}, after requiring $\text{M}_{\ell\ell\ell}<80~\text{GeV}$, the dominant background will be the SM process of $W$ boson decay, which was the subject of the previous section. Now, it has become the leading background.

\begin{figure}[t]
    \centering
    \includegraphics[width=0.48\textwidth]{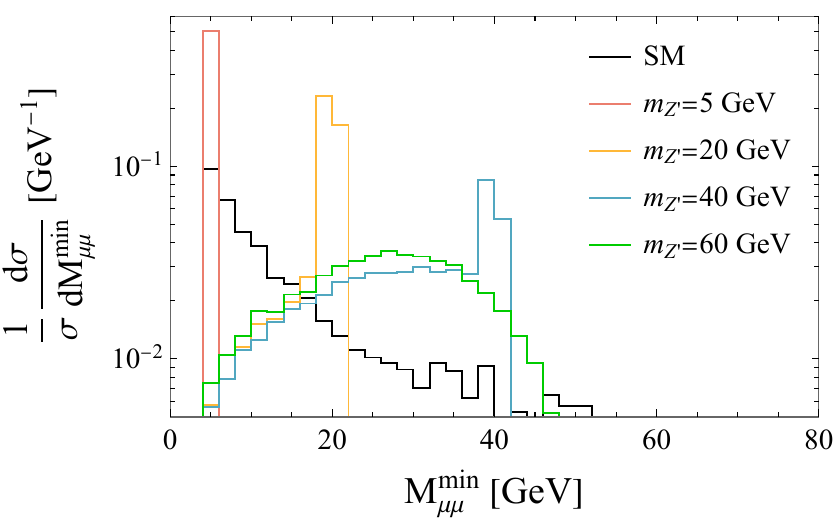}
    \caption{Reconstruction level distribution of the minimum dimuon invariant mass across different benchmarks after initial selection and $\text{M}_{\mu\mu\mu}<80~$GeV cut under HL-LHC configuration.}
    \label{fig:M_mumu}
\end{figure}

The BSM model file to produce sample events via {\tt MadGraph5\_aMC@NLO} is obtained by using {\tt FeynRules}~\cite{Alloul:2013bka}. Among the simulation work, the parton-level initial selection is the same as that in the previous section. For reconstruction-level analysis, we modify the cut-based analysis to incorporate new resonances. After requiring the invariant mass of the tri-muon system to be less than 80~GeV, we pick the pair of muons that have the minimum invariant mass $\text{M}^{\text{min}}_{\mu\mu}$ and implement an optimized mass window for each benchmark. The kinematic distribution is shown in \autoref{fig:M_mumu}.
For $W$ boson exotic decay in the SM, the $\text{M}^{\text{min}}_{\mu\mu}$ is dominantly emitted from the radiated virtual soft photon, which makes the background distribution peak at a low invariant mass regime. 
On the signal side, especially in the case of small $m_{Z'}$, the minimum-invariant-mass pair would be the pair from the $Z'$. 
For higher $m_{Z'}$ benchmarks, such an algorithm will choose the pair that does not come from $Z'$, so the $\text{M}_{\mu\mu}^{\text{min}}$ distribution gets smeared. One can take the muon pair with the highest invariant mass for a high $m_{Z'}$ benchmark or perform a lineshape fit, which will also be captured in a DNN-based analysis. We chose not to optimize further to keep the analysis simple, and this only matters for the $Z'$ mass regions near the threshold.

The minimal muon pair invariant mass range is set to be 4~GeV, aiming to avoid the $J/\psi$ production background at $3$~GeV and other sizable vector meson states. One could perform a dedicated study to optimize this regime. For $m_{Z'}$ lower than 1~GeV, asymmetric electron-positron colliders have better sensitivity~\cite{BaBar:2016sci}. Furthermore, when $m_{Z'}$ is less than 3~GeV, the decay products are too soft to pass the trigger.

\begin{figure}[t]
    \centering
    \includegraphics[width=0.445\textwidth]{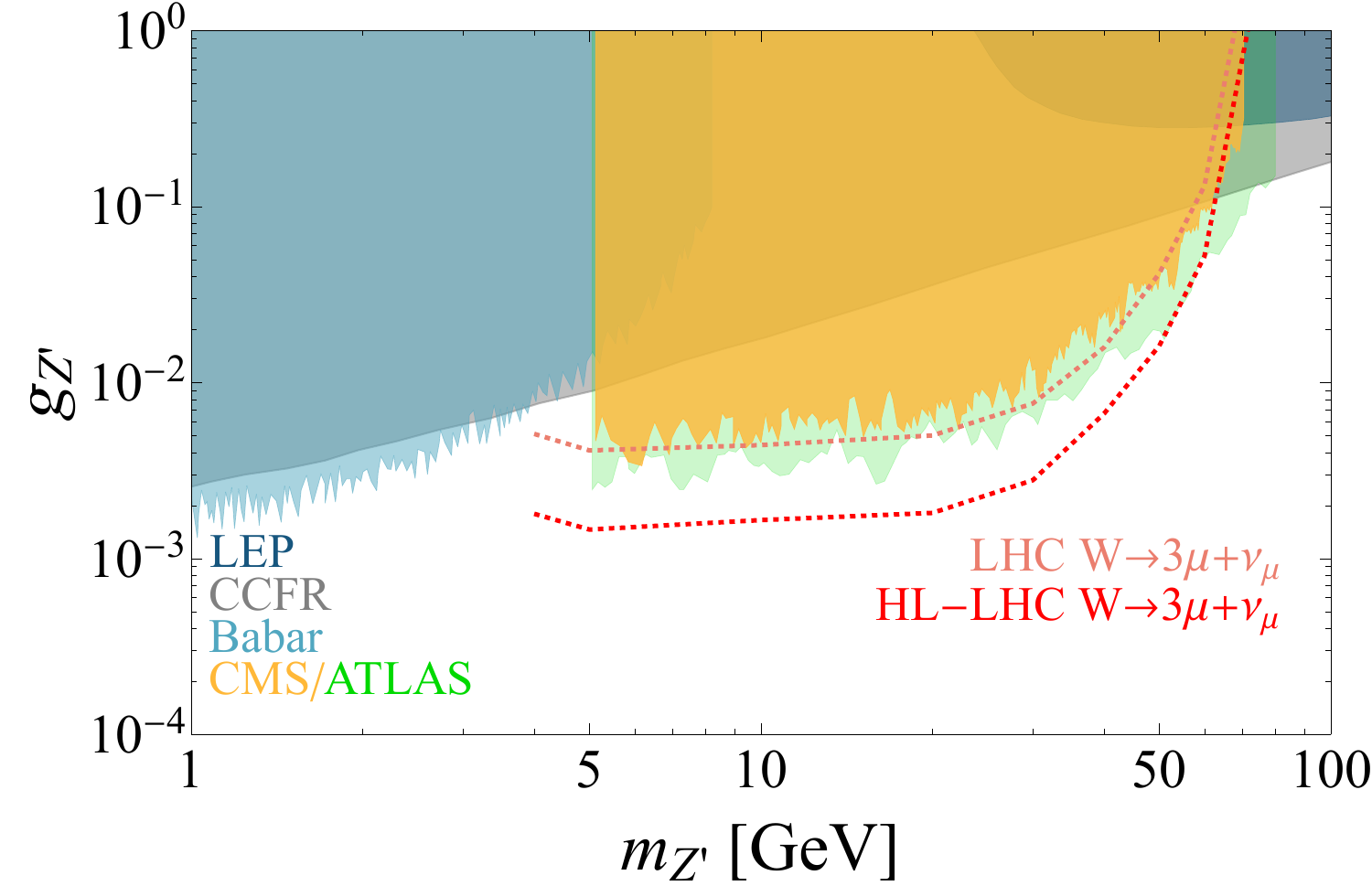}
    \caption{The 95\% C.L. projected sensitivity of $g_{Z'}$ as a function of mass $m_{Z'}$. Shaded regions are excluded by neutrino trident in CCFR~\cite{Altmannshofer:2014pba,CCFR:1991lpl}, LEP~\cite{Altmannshofer:2014cfa}, Babar~\cite{BaBar:2016sci}, CMS $Z\rightarrow 4\mu$ analysis~\cite{CMS:2018yxg}, and the recent ATLAS Run-2 $W\rightarrow 3\mu+\nu$ analysis~\cite{ATLAS:2024uvu}. 
    The dashed lines correspond to Run-2 (light-red) and HL-LHC (red) benchmarks.}
    \label{fig:Zp_sensitivity}
\end{figure}

The 95\% C.L. projected sensitivities from this $W$ exotic decay channel on coupling strength $g_{Z'}$ versus the $Z'$ mass are shown in \autoref{fig:Zp_sensitivity} with light-red and red dashed lines for our Run-2 analysis and HL-LHC, respectively. We also show the current exclusions from other existing searches. $Z\to4\mu$~\cite{CMS:2018yxg} is an established channel to look for light leptonic $Z'$. It enjoys a low SM background and is hence helpful. Our Run-2 benchmark is generally better than the $Z\to4\mu$ channel due to a higher production rate and a higher leptonic decay branching fraction of the $W$ boson. 
For the $Z'$ mass below 20 GeV, the bound for the coupling $g_{Z'}$ is almost flat since the background is mainly from SM $W$ boson exotic decay and the internal fermion propagator suppression is the same. 
Above 20 GeV, the SM background reduces, but the signal rate drops quickly due to the phase space suppression, and hence, the sensitivity drops. Close to the final stage of this study, ATLAS put out a $Z'$ search using this exotic decay channel of the $W$ boson~\cite{ATLAS:2024uvu}, which is shown as the green shaded region in \autoref{fig:Zp_sensitivity}. One can see that apart from the fluctuations from the experimental data, the projected sensitivity from our study (light-red dashed line) is highly compatible with the ATLAS search results, which indicates the robustness of our proposed analysis and hence our projections for the SM $W$ exotic decay measurement in the previous section. 
ATLAS could already carry out a few percent-level precision on this SM $W$ decay, and it is also compatible with the study's background simulation. 
Even with a slight subtlety of on-shell rate v.s. $W^*$ contributions discussed in our paper, the collaboration may want to extend their study to expand the list of the rare $W$ decay branching ratios.

In addition, we also include one interesting anomalous benchmark defined by just removing the neutrino sector from the $L_\mu-L_\tau$ model. 

\begin{figure}[th]
    \centering
    \includegraphics[width=0.45\textwidth]{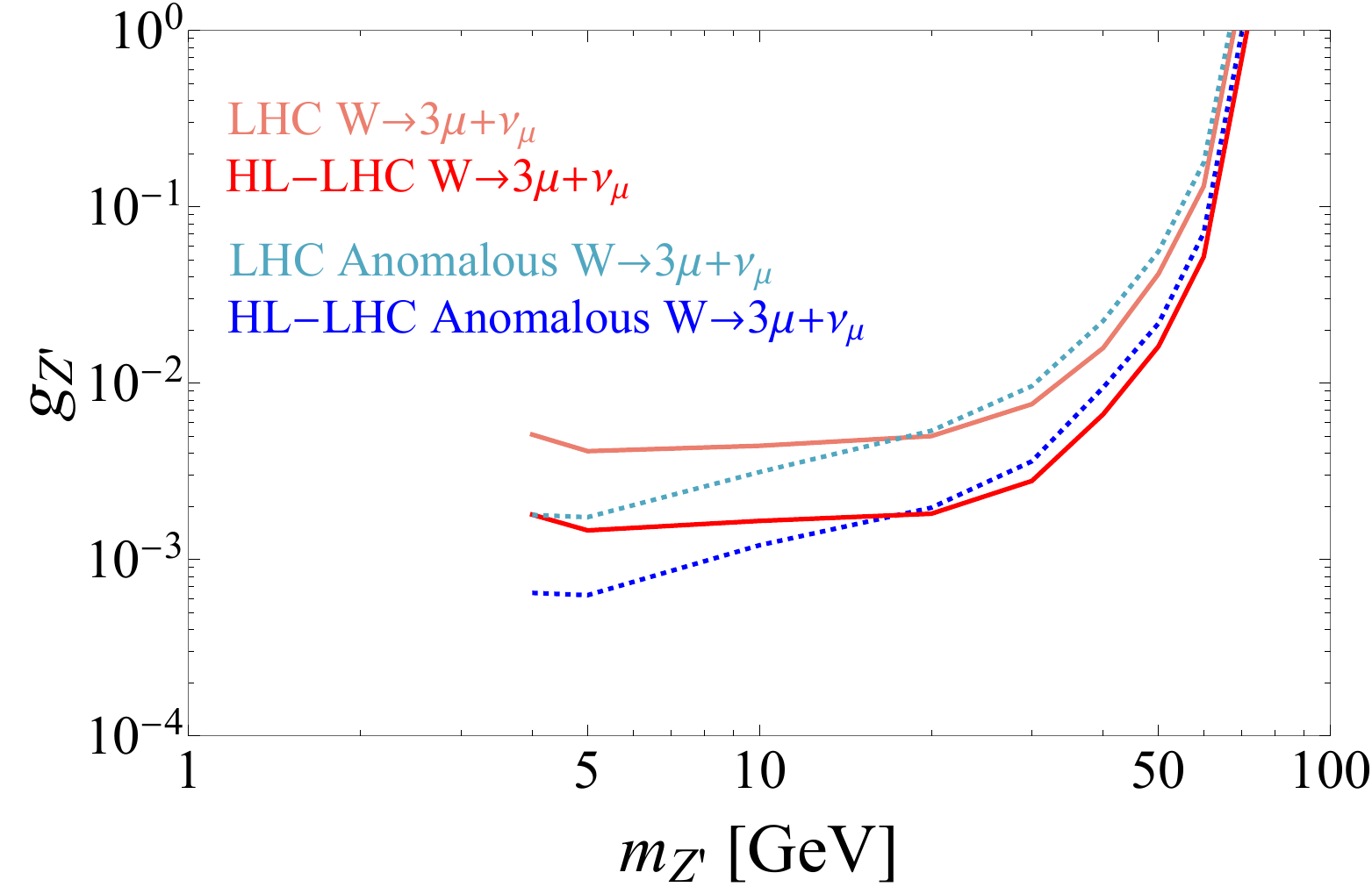}
    \caption{The 95\% C.L. projected sensitivity of $g_{Z'}$ as a function of mass $m_{Z'}$. 
    The solid lines correspond to Run-2 (light-red) and HL-LHC (red) $L_\mu-L_\tau$ benchmarks.
    The dashed lines correspond to Run-2 (light blue) and HL-LHC (blue) anomalous benchmarks.}
    \label{fig:Zp_sensitivity}
\end{figure}

We show the 95\% C.L. projected sensitivity of the anomalous $Z'$ model
\begin{align*}
    \mathcal{L} &\supset -\dfrac{1}{4} F'_{\mu\nu} F^{\prime \mu\nu} + \dfrac{1}{2} m_{Z'}^2 Z^{\prime 2}_\mu 
    + g_{Z'} Z'_\nu \brr{\bar{\mu} \gamma^\nu \mu - \bar{\tau} \gamma^\nu \tau}
\end{align*}
and compare it with the standard $L_\mu-L_\tau$ model in \autoref{fig:Zp_sensitivity}.
As is well known (see {\em e.g.} \cite{Karshenboim:2014tka,Dror:2017nsg}), 
the coupling of light $Z'$ to a non-conserved current leads to the enhancement of the longitudinal $Z'$ production by $\propto O(g^2_{Z'}E^2/m_{Z'}^2)$.  
We can see in \autoref{fig:Zp_sensitivity} that in contrast to the conventional $L_\mu-L_\tau$ model, the sensitivity in couplings gets better at smaller $Z'$, as expected.

\section{Conclusion and Outlook}
\label{sec:conclusion}

We explore the $W$ boson four-lepton decay branching ratio at LHC Run-2 and HL-LHC and propose to conduct new searches in this channel at the LHC. This new decay channel will not only add new information about the $W$ boson decays but it can also be used to search for interesting SM interference effects and directly for new physics such as $Z'$ models. This channel is statistically competitive with the $Z\rightarrow 4\ell$ channels. 

Due to the sizeable decay width of the $W$ boson, the four-lepton final states also receive contributions from other diagrams. We define the fiducial branching ratio for the invariant mass within the $[m_W - 2\Gamma_W, m_W + 2 \Gamma_W]$ range. The measurement precision from statistics can reach around $4.4\%$ for Run-2 and $0.62\%$ for HL-LHC. In addition, we exploit this decay channel to put constraints on $Z^\prime$ models. The projected sensitivity for the gauge coupling $g_{Z'}$ at HL-LHC can be improved by a factor of a few. Many new directions to further this study should be carried out in future works. For example, the flavor universality can be tested by examining the different flavor combinations for the four-lepton final states. One can also further explore the $Z$ boson propagator contribution in the exotic decay channel.

\section{Acknowledgement}
This work is supported by the Department of Energy under Grant No. DE-SC0011842 at the University of Minnesota. The data associated with the figures in this paper can be accessed via \href{https://github.com/ZhenLiuPhys/Wexotic}{Github~\faGithub}.

\appendix

\section{Branching Ratio Precision Calculation}
The fiducial branching ratio is given by:
\begin{align}
    \text{Br}^{\text{fid}}=\frac{\sigma_{\text{tot}}^{\text{fid}}-\sigma_{\text{B}}^{\text{fid}}}{\sigma_W}\times f_W
\end{align}
where $\sigma_{\text{B}}^{\text{fid}}$ is the fiducial cross section of the background, and $\sigma_{\text{tot}}^{\text{fid}}$ is the total fiducial cross section $(\sigma_{\text{B}}^{\text{fid}}+\sigma_{\text{S}}^{\text{fid}})$.

The statistical uncertainty is computed via error propagation and the Poisson distribution $Poi(N)$ with standard deviation $\sqrt{N}$:
\begin{align}
    \delta \text{Br}^{\text{fid}}=\sqrt{\left( \frac{\partial\text{Br}^{\text{fid}}}{\partial \sigma_i}\delta \sigma_i\right)^2} \quad \text{with}~ \sigma_i=\sigma_{\text{tot}}^{\text{fid}},\sigma_{\text{B}}^{\text{fid}},\sigma_W
\end{align}
and
\begin{align}
    \frac{\delta \text{Br}^{\text{fid}}}{ \text{Br}^{\text{fid}}}=\sqrt{\frac{(\delta\sigma_{\text{tot}}^{\text{fid}})^2+(\delta\sigma_{\text{B}}^{\text{fid}})^2+\left(\frac{\sigma_{\text{tot}}^{\text{fid}}-\sigma_{\text{B}}^{\text{fid}}}{\sigma_W}\delta\sigma_W\right)^2}
    {(\sigma_{\text{tot}}^{\text{fid}}-\sigma_{\text{B}}^{\text{fid}})^2}}.
\end{align}
The statistical uncertainty of the cross section $\delta\sigma_i$ can be computed from:
\begin{align}
    \delta\sigma_i\cdot L&=\delta N_i=\sqrt{N_i}~\longrightarrow~ \delta\sigma_i=\sqrt{\frac{\sigma_i}{L}},
\end{align}
where the integrated luminosities $L$ are $3~\text{ab}^{-1}$ and $150~\text{fb}^{-1}$ for the HL-LHC and Run-2 benchmarks, respectively.

\section{DNN Architecture}
\label{sec:DNN}

Each event consists of three leptons and at most two jets, both of which can be $b$-jets. 

Given the particle information, an extended version of the particle flow network (PFN), namely PFN-ID, is used. This approach incorporates PID attributes for each particle in addition to their momentum or jet constituents, as detailed in Ref.~\cite{Komiske:2018cqr}. To ensure the optimal input format for the network, the initial PID information derived from the Particle Data Group (PDG) database about leptons, and the $b$-tag values for jets, are systematically mapped onto the real number continuum within the interval [-1, 1], preserving the uniqueness of the mapping process, as described in Ref.~\cite{Komiske:2018cqr}.

Phi sizes of (64, 64, 64) and F sizes of (128, 128, 128) with 100 epochs are found to be the best configuration. To prevent overfitting, the training protocol is ceased immediately when there is a discernible divergence in the training accuracy and validation accuracy over a span of five epochs.

\bibliography{references}

\end{document}